\documentclass[aps, preprint, superscriptaddress]{revtex4-2}
\usepackage{graphicx}
\usepackage{makecell, multirow}
\usepackage{color, bm, soul, xcolor}
\usepackage{amsmath, amssymb}
\usepackage{gensymb,textcomp}
\usepackage{geometry,textcase}
\usepackage{braket}
\usepackage[colorlinks=true, linkcolor=blue, citecolor=blue, urlcolor=blue]{hyperref}
\usepackage{graphicx} 
\usepackage[english]{babel}
\newcommand{\eg}{{\em e.\,g.}}

\begin{document}

\title{Dipolar Nematic State in Relaxor Ferroelectrics}
\author{Yuan-Jinsheng Liu}
\affiliation{Department of Physics, School of Science, Westlake University, Hangzhou, Zhejiang 310030, China}

\author{Tyler C. Sterling}
\affiliation{Department of Physics, University of Colorado, Boulder, Boulder, CO 80303, USA}

\author{Shi Liu}
\email{liushi@westlake.edu.cn}
\affiliation{Department of Physics, School of Science, Westlake University, Hangzhou, Zhejiang 310030, China}
\affiliation{Institute of Natural Sciences, Westlake Institute for Advanced Study, Hangzhou, Zhejiang 310024, China}

\begin{abstract}
Relaxor ferroelectrics exhibit exceptional dielectric and electromechanical 
properties, yet their microscopic origins remain elusive due to the interplay of  hierarchical polar structures and chemical complexity. While models based on polar nanoregions or nanodomains offer  valuable phenomenological insights, they often lack the first-principles predictive capability necessary for quantitatively describing functional properties such as piezoelectric coefficients. Here, we use large-scale molecular dynamics simulations, enabled by a universal first-principles-based machine-learning interatomic potential, to investigate atomic-scale polar dynamics in canonical Pb-, Bi-, and Ba-based relaxors. Across all systems, we uncover a universal dipolar nematic state, characterized by long-range orientational order of local polarizations without local alignment, challenging conventional polar cluster-based paradigms. We introduce a universal order parameter, derived from the skewness of the distributions of the local polarization autocorrelation functions, that captures the thermal evolution of both lead-based and lead-free systems within a single master curve. This nematic order, and its robust structural memory under electric field cycling, underpins key relaxor phenomena, including diffuse phase transition, frequency-dependent dielectric dispersion,  and reversible giant piezoelectricity. Our findings establish a unified microscopic framework for relaxors and present a broadly applicable statistical approach to understanding complex disordered materials.
\end{abstract}

\maketitle

\clearpage
\section{Introduction}
Relaxor ferroelectrics represent a class of functional materials with high susceptibilities to external electrical, mechanical, and thermal perturbations. Their exceptional properties, such as colossal dielectric permittivity, giant piezoelectricity, and strong pyroelectric effects, make them indispensable for next-generation actuators, sensors, and advanced energy technologies~\cite{Park97p1804,Pandya18p432,Kutnjak06p956,Li25peadn4926}.  At a fundamental level, their behavior is governed by a delicate interplay between hierarchical polar structures and underlying chemical disorder. Understanding how robust, emergent properties arise from a complex landscape of competing interactions makes the study of relaxors not only a long-standing pursuit in materials science but also a vital endeavor in the broader context of complexity physics.

The structural foundation for most high-performance relaxor ferroelectrics is the perovskite-type $AB$O$_3$ lattice, where the $A$-site typically hosts larger cations (e.g., Pb$^{2+}$) and the $B$-site accommodates smaller transition metals (e.g., Ti$^{4+}$ and Nb$^{5+}$). Understanding these materials remains a challenge, largely due to their compositional complexity and the absence of a unified theoretical framework. 
Modern relaxors often involve complex solid solutions, such as Pb(Mg$_{1/3}$Nb$_{2/3}$)O$_3$–PbTiO$_3$ (PMN–PT), particularly near morphotropic phase boundary (MPB) compositions \cite{Park97p1804,Li19p264,Rojac23p12}, where multiple cations occupy the same Wyckoff position, introducing spatially varying chemical disorder.
Even in extensively studied lead-based systems, theoretical models diverge in their interpretations of how polar order and chemical disorder interact to determine functional properties. Polar cluster-based theories suggest that nanoscale polar entities form the fundamental building blocks. Among these, the polar nanoregion (PNR) model proposes polar clusters embedded in a nonpolar matrix \cite{Burns73p423,Burns83p853}, while the polar nanodomain (PND) model places them in a weakly polar matrix \cite{Takenaka17p391,Kim22p1502}.
Other frameworks offer alternative perspectives. The random-field model attributes the disruption of long-range ferroelectric order to static chemical disorder \cite{Westphal92p847,Phelan14p1754}; 
the polarization rotation model explains the giant piezoelectric response through a monoclinic phase-mediated polarization rotation mechanism or the intrinsic softness of the rhombohedral and orthorhombic phases \cite{Fu00p281,Kutnjak06p956,Noheda02p27}; the adaptive phase model interprets nanoscale structures as mixtures of strain-stabilized phases \cite{Jin03p3629,Jin03p197601}.
While each model captures certain aspects, none offer predictive, first-principles-based descriptions of key properties, such as giant piezoelectric coefficients, directly from chemical composition.

This theoretical gap becomes more pronounced when considering diverse ways in which chemical disorder and ferroelectric activity are distributed across the $A$ and $B$ sites in different chemical families. In PMN-PT, ferroelectric activity arises from both the $A$-site Pb$^{2+}$ and $B$-site Ti$^{4+}$ cations, with disorder present on the $B$-site. In contrast, the lead-free relaxor (Bi,~Na)TiO$_3$ exhibits $A$-site disorder, where Bi$^{3+}$ and Na$^+$ share the $A$-site, and both Bi$^{3+}$ and Ti$^{4+}$ contribute to ferroelectricity. In Ba(Zr,~Ti)O$_3$, ferroelectric-active cations coexist with chemical disorder on the $B$-site.
These variations illustrate that relaxor behavior can emerge from different combinations of site-specific disorder and ferroelectric mechanisms. A framework that can account for this diversity, spanning both lead-based and lead-free systems, is lacking.

To address these long-standing challenges, we employ large-scale molecular dynamics (MD) simulations, enabled by a first-principles-based universal interatomic potential \cite{Wu23p180804}, to investigate the atomic-scale structural dynamics across a chemically diverse set of relaxor families, including Pb-, Bi-, and Ba-based systems. Here, we reveal a universal organizing principle, which we identify as a dipolar nematic state. This state is defined by the presence of long-range orientational order among unit-cell-resolved local polarizations. Surprisingly, this order persists without requiring strict local parallel alignment of local polarizations, in sharp contrast to the collinear dipole alignments typically assumed in polar cluster-based relaxor models.

To quantitatively describe this complex state, we introduce a statistical approach that yields a universal order parameter, derived from the skewness of the distribution of local polarization autocorrelation functions. This single metric successfully collapses the thermal evolution of all three relaxor systems onto a unified master curve. It also captures the relaxor's intrinsic ``structural memory”, the ability of the nematic ground state to recover after the removal of an external stimulus, which we demonstrate as the microscopic origin of the giant and reversible electromechanical response.
The dipolar nematic model offers a conceptual framework for understanding both lead-based and lead-free relaxors, serving as a predictive foundation for designing high-performance relaxors.

\section{Results}
\subsection{MD simulations of Pb-based relaxors}
Extensive efforts have been made to uncover the atomic-scale origins of functional properties of lead-based relaxors represented by PMN-PT \cite{Li16p13807,Takenaka17p391,Eremenko19p2728,Kumar20p62,Kim22p1502,Kim25p478}. Establishing a direct connection between structural dynamics and the macroscopic functional responses  requires efficient computational methods capable of large-scale simulations. While previous MD studies have successfully reproduced diffuse scattering patterns of PMN-PT, they were performed under constant-volume constraints and did not capture key relaxor behaviors, including the giant electromechanical response \cite{Takenaka17p391}. This challenge is even greater in the more chemically complex ternary relaxor Pb(In$_{1/2}$Nb$_{1/2}$)O$_3$\allowbreak{}–\allowbreak{}Pb(Mg$_{1/3}$Nb$_{2/3}$)O$_3$\allowbreak{}–\allowbreak{}PbTiO$_3$ (PIN–PMN–PT), which is of particular interest for its large piezoelectric coefficients and high Curie temperature \cite{Zhang08p064106,Liu09p074112}. Yet, its extreme compositional complexity has long hindered a definitive atomistic understanding. 
Here, we present comprehensive MD simulations of PIN–PMN–PT, directly bridging its microstructural complexity with emergent functional behavior.

\begin{figure}[t]
\includegraphics[width=0.9\textwidth]{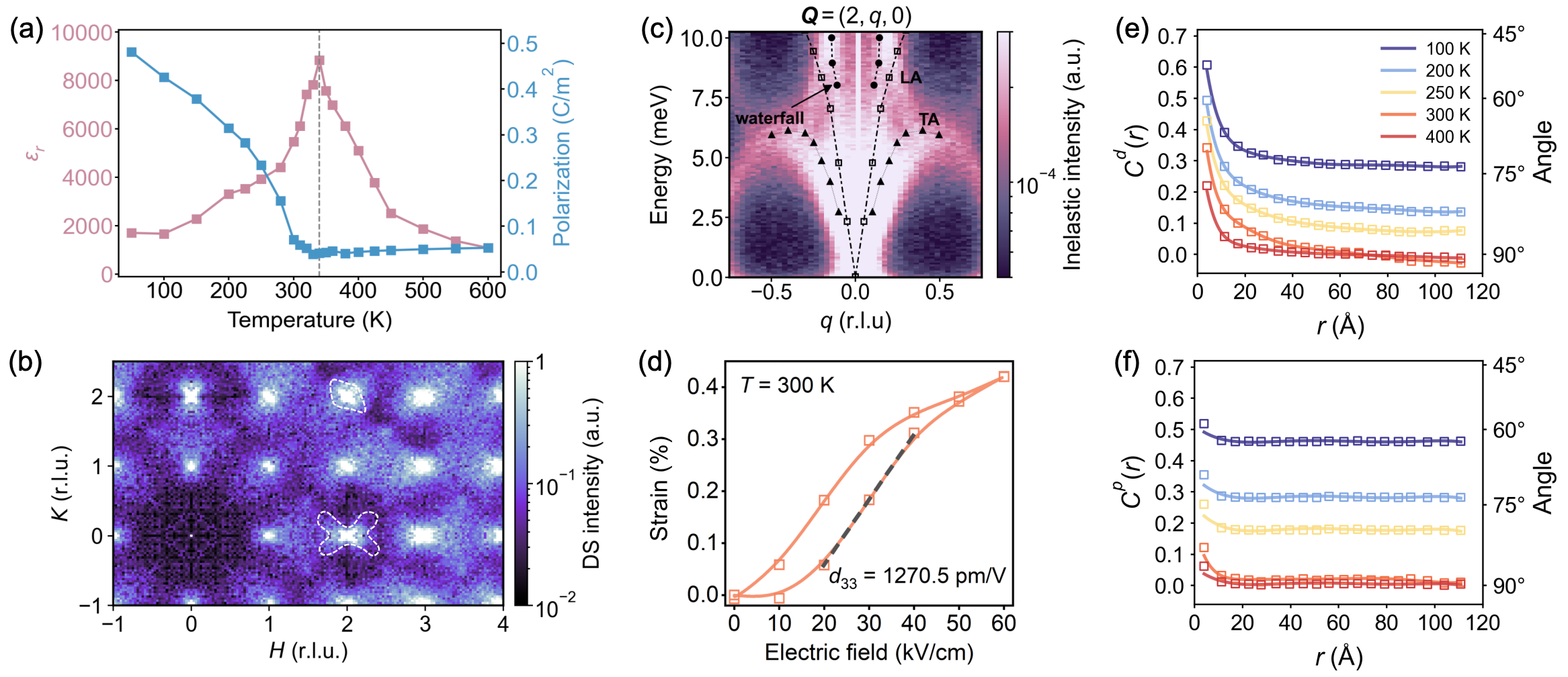}
\caption{\textbf{MD simulations of the lead-based relaxor 0.24PIN--0.42PMN--0.34PT.}
\textbf{a}, Simulated static dielectric permittivity ($\epsilon_r$) and total polarization as a function of temperature. The relaxor exhibits a broad dielectric peak at $T_\mathrm{m} \approx 340$~K and a gradual polarization decay, characteristic of a diffuse phase transition.
\textbf{b}, Simulated elastic neutron diffuse scattering patterns in the $(H, K, 0)$ plane at 300~K. The patterns reproduce the experimentally observed butterfly-shaped and ellipsoidal distributions  around the (200) and (220) Bragg peaks, respectively~\cite{Li18p345,Bosak11p117,Krogstad18p718}.
\textbf{c}, Simulated inelastic neutron scattering spectra, compared to phonon peak positions measured in experiments (markers)~\cite{Li18p345}. Empty markers
indicate branches that are weak in the plotted zone; filled markers are branches that are strong. The waterfall feature along the [100] direction is clear. 
\textbf{d}, Strain-electric field hysteresis loop, yielding a giant effective piezoelectric coefficient ($d_{33}$) of $>1200$~pC/N.
Ensemble-averaged spatial correlation function of \textbf{e}, Pb displacements $C^d(r)$, and \textbf{f}, local polarization $C^p(r)$, at various temperatures. Corresponding angles are shown, with 90\degree~indicating no correlation. The decay to zero at large distances indicates the loss of long-range polar order, while saturation at non-zero values at lower temperatures reflects a ferroelectric phase.}
\label{fig:lead}
\end{figure}

The UniPero model \cite{Wu23p180804}, trained exclusively on first-principles data, enables large-scale MD simulations of the 0.24PIN–0.42PMN–0.34PT composition, a MPB relaxor, under constant-temperature, constant-pressure ($NPT$) conditions. These simulations successfully reproduce experimental hallmarks of relaxor behavior.
First, they capture the gradual reduction of the total polarization with increasing temperature, as well as the broad peak in dielectric permittivity characteristic of relaxors ($T_m \approx 340$~K, see Fig.~\ref{fig:lead}a), different from the abrupt transition typical of simple ferroelectrics like PbTiO$_3$. Second, neutron scattering spectra computed from MD configurations at 300~K (slightly below $T_m$) reproduce key experimental signatures. These include the anisotropic diffuse scattering (DS) patterns, specifically, the butterfly-shaped and ellipsoidal intensity distributions around the (200) and (220) Bragg peaks, respectively (Fig.~\ref{fig:lead}b) ~\cite{Li18p345,Bosak11p117,Krogstad18p718}. 
In comparison, the anisotropic DS is significantly reduced at 500~K (Fig.~S1).
Decomposition of the DS intensity by atomic sublattice (see Supplementary Sec.~I) reveals that the the butterfly-shaped and ellipsoidal features originate primarily from the Pb and O atoms, while the $B$-site sublattice contributes a large, nearly temperature-independent background (Fig.~S2). A quantitative analysis of DS intensity along the $(2+q, 2-q, 0)$ direction around the (220) Bragg peak shows a Lorentzian decay with intensity scaling as $1/q^2$ (Fig.~S3), consistent with experimental results \cite{Chetverikov02ps989,Xu04p064112}.
Furthermore, our simulated inelastic neutron scattering spectra (Figs.~S4-S5) also agree well with the measured phonon dispersion~\cite{Li18p345}, notably the waterfall effect. As shown in Fig.~\ref{fig:lead}c, along $\bm{Q} = (2,q,0)$, the waterfall and transverse acoustic (TA) modes are strong, while the longitudinal acoustic (LA) mode is weak, matching experimental observations~\cite{Li18p345}; a similar waterfall feature is also seen near the (220) peak (Fig.~S4).

Lastly, by directly simulating the strain-electric field hysteresis at 300 K, we obtain an effective piezoelectric coefficient of $>1200$~pC/N (Fig.~\ref{fig:lead}d). Reproducing these relaxor signatures, including the diffuse phase transition, anisotropic diffuse scattering, waterfall effects, and giant piezoelectricity, strongly validates our approach. To our knowledge, this represents the first first-principles-based model to achieve such comprehensive agreement with experiments for a complex ternary relaxor. 

\subsection{Ensemble-averaged structural features in Pb-based relaxors}

To probe the system's polar microstructure, we first compute the ensemble-averaged, time-delayed orientational correlation functions, $C^d(r, \tau)$ and $C^{p}(r, \tau)$ (see Methods), which quantify the temporal and spatial correlations of the $A$-site Pb displacements ($d$) and the local polarization ($p$, defined as the electric dipole moment per unit cell volume), respectively. Surprisingly, both correlation functions show minimal dependence on $\tau$ for $\tau > 2.0$ ps (Fig.~S6), indicating that the relative orientations (angles) between local displacement or polarization vector pairs are quasi-static, remaining effectively frozen over the timescales probed. This allows us to simplify the analysis by averaging over the time delay: $C^d(r) \equiv \langle C^{d}(r, \tau) \rangle_\tau$ and $C^{p}(r) \equiv \langle C^{p}(r, \tau) \rangle_\tau$.

The spatial correlation of the $A$-site displacements, $C^{d}(r)$, as a function of temperature is shown in Fig.~\ref{fig:lead}e. Across all temperatures, $C^d(r)$ decays sharply at short distances ($r < 18$~\AA). Importantly, the data reveal several notable deviations from classical polar cluster-based paradigms. Even for nearest-neighbor Pb ions ($r_{nn} \approx 4$~\AA), the correlation is moderate. For example, $C^d(r_{nn}) \approx 0.6$ at 100~K, corresponding to an average angle of $\approx53$\degree~between neighboring displacement vectors. This weak local alignment contrasts with the PNR and PND models, which assume strongly aligned local polar displacements within polar clusters. At low temperatures (\eg, 100--250~K) well below $T_m$, $C^d(r)$ saturates to a finite, non-zero value at long range, consistent with the emergence of global order and macroscopic polarization shown in Fig.~\ref{fig:lead}a. Near $T_m$, as seen at 300 K, $C^d(r)$ decays more gradually but ultimately vanishes at large $r$, suggesting that only short- to intermediate-range correlations survive. At 400~K (above \( T_m \)), \( C^d(r) \) starts from $\approx 0.2$ (average angle $\approx79$\degree) and decays rapidly to zero. This reflects weak short-range correlations and a loss of extended polar order.

The local polarization, which includes  contributions from both $A$- and $B$-site cation displacements, provide a more comprehensive measure of local symmetry breaking (see Methods). Notably, as shown in Fig.~\ref{fig:lead}f, $C^{p}(r)$ remains nearly flat across all spatial separations. At 100~K, $C^{p}(r)$ begins at $\approx0.52$ for nearest neighbors and decreases only slightly to around 0.47 at long range. The corresponding $\approx58^\circ$ average angle between neighboring local polarization vectors again indicates modest local alignment, similar to $C^{d}(r_{nn})$ at the same temperature. The absence of sharp short-range decay in $C^{p}(r)$ suggests that including $B$-site contributions smooths out polar inhomogeneities. 
The flat spatial profile of $C^{p}(r)$, combined with its finite value, reflects a counterintuitive system-wide polar coherence despite weak local alignment. This behavior contrasts with fragmented and randomly oriented polar domains, which would drive $C^{p}(r)$ to zero at long range. 
At higher temperatures such as 300 K and 400 K, $C^p(r)$ decays rapidly to zero,  despite $T_m$ lying between these temperatures. This suggests that long-range polar order remains suppressed below $T_m$. Such decoupling between the structural transition (associated with macroscopic polarization) and the dielectric peak is a defining feature of relaxor behavior. 

Overall, these temperature-dependent ensemble-averaged orientational correlation functions corroborate the temperature dependence of the macroscopic polarization (Fig.~\ref{fig:lead}a). Moreover, they reveal two unconventional features. First, the absence of strong local polar alignment at all temperatures rules out the presence of well-defined polar clusters. Second, the spatially uniform and time-delay-independent behavior of $C^p(r)$ suggests a quasi-static, distance-independent correlations among local polarizations, rather than dynamic, constantly reorienting polar domains.

\subsection{Dipolar nematic state in Pb-based relaxors}
\begin{figure}[t]
\includegraphics[width=0.85\textwidth]{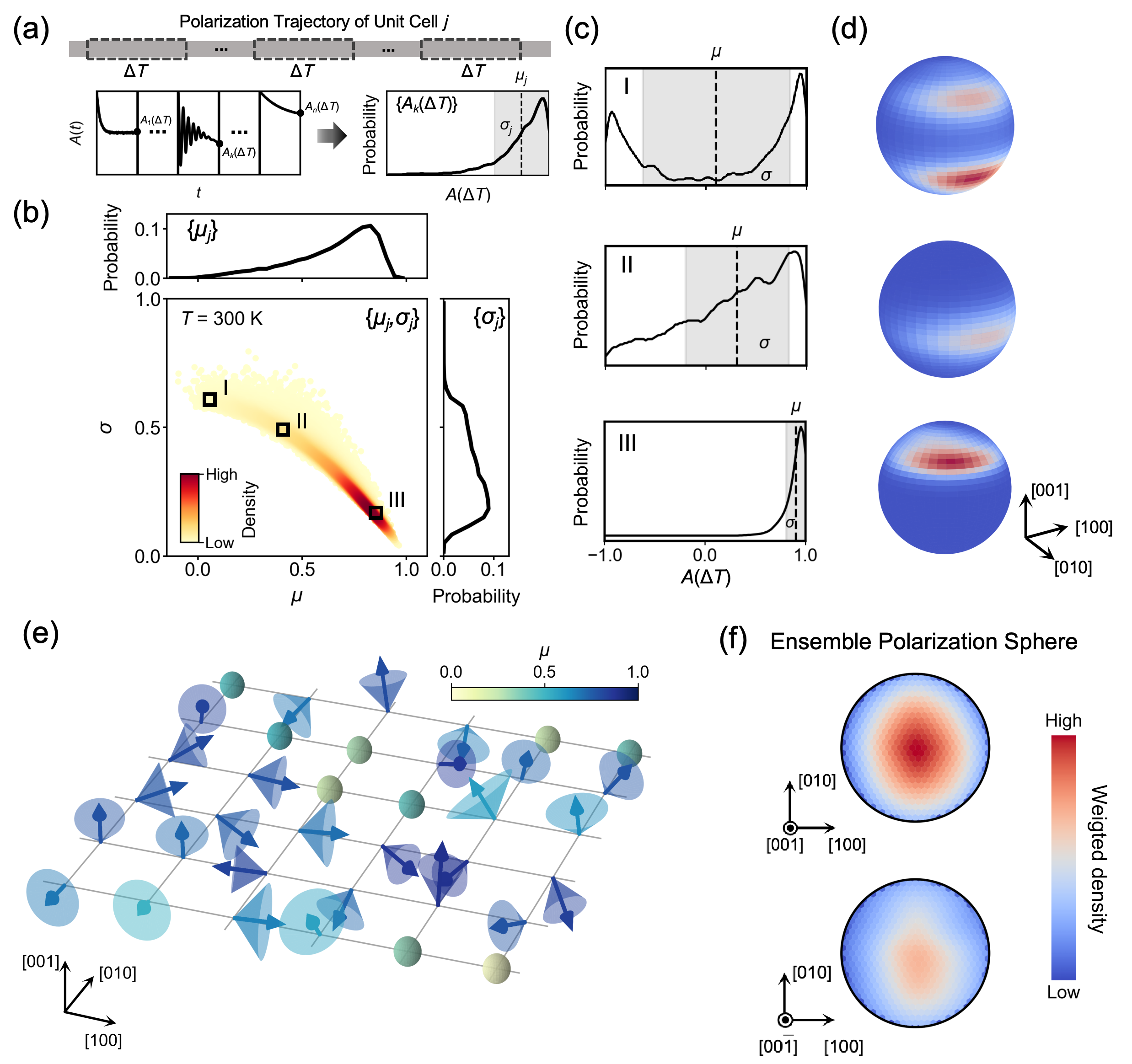}
\caption{\textbf{Dipolar nematic state in the Pb-based relaxor 0.24PIN--0.42PMN--0.34PT.}
\textbf{a}, Schematic of the sliding-window analysis used to 
construct the distribution $\{ A_k({\Delta T}) \}$ from the local polarization autocorrelation function, based on the polarization trajectory of unit cell $j$. This distribution yields the mean ($\mu_j$, dashed line) and standard deviation ($\sigma_j$, shaded region), which together  characterize the polarization dynamics of unit cell $j$.
\textbf{b}, Scatter plot of the $\{\mu_j$, $\sigma_j\}$ distributions for all unit cells at 300~K, overlaid with marginal distributions. The distribution is skewed towards high $\mu$ and low $\sigma$.
\textbf{c}, Classification of unit cells into three dynamic types based on their $\{A_k({\Delta T})\}$ distributions.
\textbf{d}, Representative polarization trajectories for each type, projected onto the unit sphere: Type-I (stochastic), Type-II (diffuse), and Type-III (persistent).
\textbf{e}, Arrow-cone visualization of a representative region, showing that local polarizations fluctuate around preferred  directions but lack parallel alignment (polar cluster). Color indicates $\mu_j$.
\textbf{f}, Ensemble polarization sphere, constructed from instantaneous local polarization vectors of all unit cells over a 1 ns MD trajectory at 300~K. The emergence of a single dominant axis confirms long-range orientational order.}
\label{fig:acf}
\end{figure}

Ensemble-averaged correlation functions 
challenge polar cluster-based relaxor models.
To further quantify the spatiotemporal evolution of the local polarization, we perform a sliding-window analysis of the time-dependent polarization vectors for each unit cell (Fig.~\ref{fig:acf}a). The polarization trajectory of a given unit cell $j$ is divided into overlapping intervals of fixed duration $\Delta T$ (\eg, 200~ps). For each interval $k$, we compute the local polarization autocorrelation function (ACF), $A_k(t)$, and extract a single scalar value at $t=\Delta T$, denoted $a_k \equiv A_k(t = \Delta T)$. Sliding the window across the trajectory yields a distribution of such values, $\{a_k\}$. From this, we compute the mean $\mu_j$ and standard deviation $\sigma_j$ for unit cell $j$, reflecting the average persistence and temporal variability of its local polarization, respectively.
Specifically, a larger $\mu_j$ indicates the local polarization of unit cell $j$ remains  closely aligned with its original direction after time $\Delta T$, signaling smaller temporal fluctuations; a larger $\sigma_j$ typically means the polarization undergoes hopping among multiple directions. 
Extending this analysis to all unit cells yields system-wide distributions of $\{\mu_j\}$ and $\{\sigma_j\}$, providing a statistical perspective on the spatiotemporal behavior of the polar structure (see additional details in Methods).

Figure~\ref{fig:acf}b presents a scatter plot of $(\mu_j, \sigma_j)$ values for all unit cells at 300~K, where each point represents a single unit cell and color indicates point density. 
The distribution is strongly skewed toward high $\mu$ and low $\sigma$, with a pronounced peak near $\mu \approx 0.7$ and most $\sigma$ values below 0.5.  
To interpret this distribution, we classify unit cells into three dynamic types based on their $(\mu_j, \sigma_j)$ values, as shown in Fig.~\ref{fig:acf}c. Type-III cells (high $\mu$, low $\sigma$) display strong directional memory with limited variability. For example, the polarization trajectory of a representative Type-III cell, projected onto a unit sphere over 1 ns (Fig.~\ref{fig:acf}d, bottom), reveals confined rotational motion about a single preferred direction.
In contrast, Type-II cells (intermediate $\mu$, moderate $\sigma$) exhibit broader distributions of polarization directions, reflecting more diffuse orientational dynamics (Fig.~\ref{fig:acf}d, middle). Type-I cells (low $\mu$, high $\sigma$) show bimodal distributions, consistent with stochastic switching between two distinct orientations (Fig.~\ref{fig:acf}d, top). These dynamic subtypes reflect a broad spectrum of local polarization behaviors that are averaged out in conventional ensemble analyses. As evident from the density distribution in Fig.~\ref{fig:acf}b, the majority of unit cells fall into the Type-III category that maintains relatively persistent polarization directions with moderate temporal fluctuations.

To further examine the spatial organization of local polarizations, we visualize each unit cell’s polarization as a cone. The cone’s direction corresponds to the time-averaged polarization vector, while its angular spread is scaled by $\mu_j$ to reflect the magnitude of temporal fluctuations. 
Specifically, Type-III cells with high $\mu_j$ appear as narrow cones, indicating stable polarization directions; Type-I and Type-II cells with low $\mu_j< 0.5$ are rendered as nearly spherical markers to enhance visual clarity. 
Figure~\ref{fig:acf}e displays a representative region of the system at 300~K, illustrating that while local polarizations fluctuate around preferred orientations, their relative angles remain both dynamically and statistically constrained. This aligns with the behavior of  $C^p(r)$, showing that neighboring polarization vectors seldom adopt parallel alignment.
To assess global orientational order, we project all instantaneous local polarization vectors of all unit cells of a 1~ns trajectory onto the surface of a unit sphere to construct the ensemble polarization sphere (Fig.~\ref{fig:acf}f). The color map on the sphere encodes both the directional probability density and the magnitude of polarization.
This construction reveals a clear anisotropy, with local polarization vectors preferentially aligned along the [001] crystallographic axis. Although both [001] and [00$\bar{1}$] directions are populated, the distribution is slightly biased toward [001], resulting in partial cancellation and a weak global polarization.

Taken together, these results indicate that at 300~K, just below \( T_m \), the system lacks domain-level polar alignment, as seen in the polarization cone plot (Fig.~\ref{fig:acf}e), and exhibits extremely weak long-range polar order, as demonstrated by the decay of \( C^p(r) \) to zero at large distances (Fig.~\ref{fig:lead}f). 
Despite this, the system retains persistent long-range orientational correlations, revealed by the anisotropic distribution on the ensemble polarization sphere featuring high intensity along the [001] crystallographic axis  (Fig.~\ref{fig:acf}f). This coexistence of local polar disorder and sustained orientational coherence is indicative of a \textit{dipolar nematic state}, analogous to the nematic phase in liquid crystals, where molecules lack positional order but align collectively along a preferred axis \cite{khoo22liquid}. 
Our calculations of the nematic order parameter 
$S$ in 0.24–0.42PMN–0.34PT reveal that it exhibits a temperature dependence similar to that of the macroscopic polarization (see Fig.~S7).
We note that previously reported domain-like features in relaxor systems may be artifacts of ensemble-averaged configurations in MD simulations (see Fig.~S8), potentially obscuring  the underlying dipolar nematic state.
Upon further cooling, the emergence of macroscopic polarization signals a transition into a \textit{ferroelectric nematic state} (see Fig.~S11), in which both long-range orientational and polar order are established. In this regime, evident at 250~K in Fig.~\ref{fig:lead}f, \( C^p(r) \) saturates at a finite value at large distances, consistent with the substantial macroscopic polarization observed in Fig.~\ref{fig:lead}a. It is remarkable that only through the integrated analysis of macroscopic polarization, the spatial decay of \( C^p(r) \), the local polarization cone plot, and the ensemble polarization sphere can the intricate and highly correlated nature of the system’s dynamic structure become apparent.

\subsection{Dipolar nematic state in Bi-based relaxors}

The (Bi$_{0.5}$Na$_{0.5}$)TiO$_3$ (BNT) solid solution is a widely studied lead-free relaxor system, yet its temperature-dependent structural evolution remains a long-standing subject of debate \cite{Shvartsman11p1,Zhou21p100836}. Dielectric permittivity measurements have revealed two notable features: a broad maximum at $T_m$ and a frequency-dependent hump near the depolarization temperature $T_d$ \cite{Sakata74p347,Hiruma09p084112,Rao13p224103}. The microscopic nature of the intermediate phase between \( T_d \) and \( T_m \) remains controversial, with interpretations ranging from antiferroelectric ordering \cite{Pronin80p395,Takenaka91p2236,Tu94p11550}, relaxor behavior \cite{Shvartsman11p1,Tu94p11550}, to phase coexistence \cite{Suchanicz88p107,Suchanicz95p249,Rao13p060102}.

\begin{figure}[t]
\includegraphics[width=0.9\textwidth]{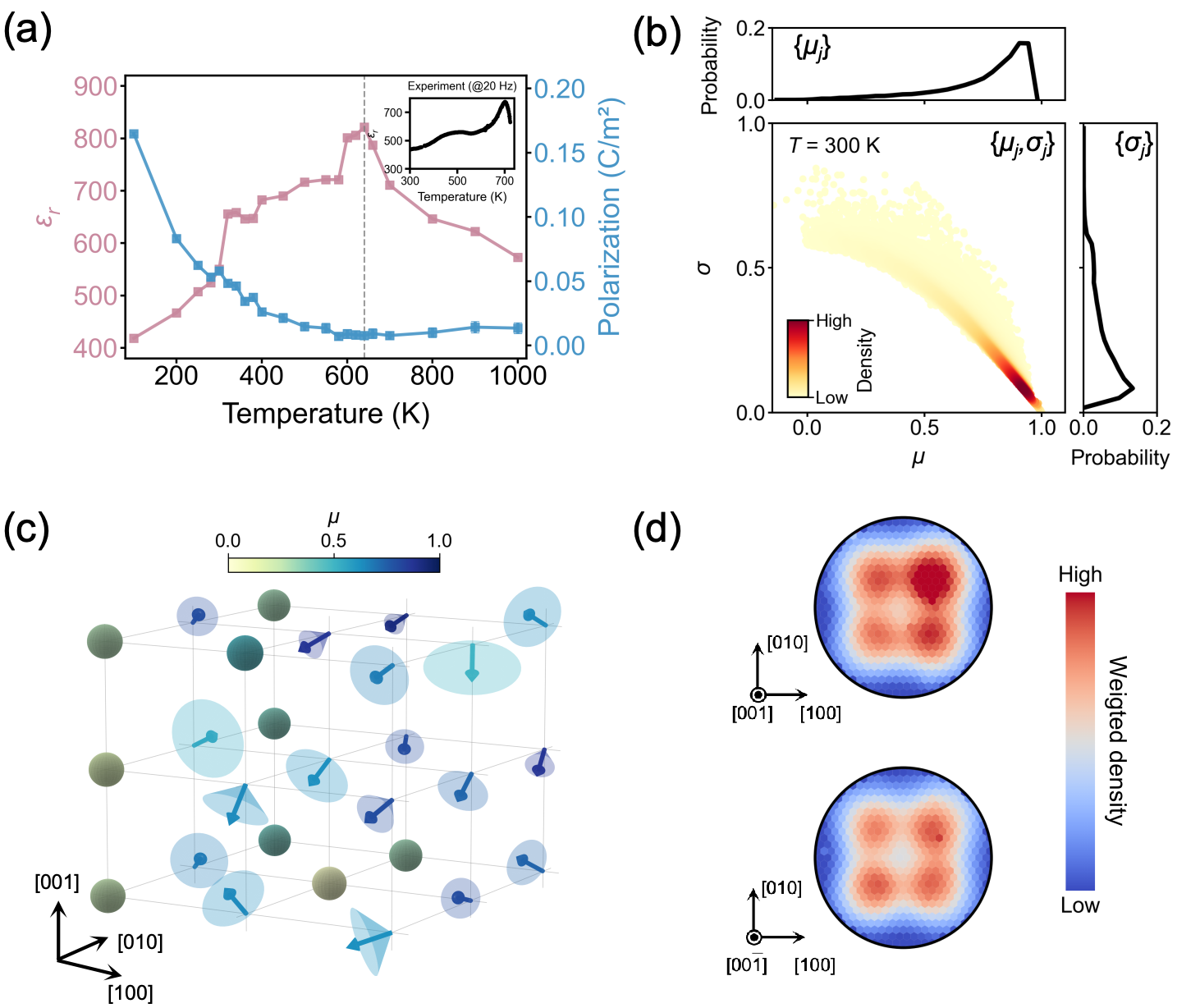}
\caption{
\textbf{Multipolar nematic state in the Bi-based relaxor Bi$_{0.5}$Na$_{0.5}$TiO$_3$.}  
\textbf{a}, Simulated static dielectric permittivity and total polarization of Bi$_{0.5}$Na$_{0.5}$TiO$_3$ as functions of temperature. The simulation reproduces key experimental anomalies (inset): a dielectric hump at \( T_{d} \approx 320 \)~K and a broad peak at \( T_{m} \approx 640 \)~K.  
\textbf{b}, Scatter plot of the $\{\mu_j$, $\sigma_j\}$ distributions at 300~K, showing a dominance of unit cells with Type-III dynamics, characterized by high \( \mu \) and low \( \sigma \).  
\textbf{c}, Arrow-cone visualization of a representative region, illustrating the absence of local parallel alignment.  
\textbf{d}, Ensemble polarization sphere showing eight distinct, symmetry-related preferred orientations, in contrast to the single orientation observed in PIN-PMN-PT. This coexistence of multiple orientational axes and local disorder defines the multipolar nematic state.
}
\label{fig:BNT}
\end{figure}

Our MD simulations accurately capture the complex dielectric response of the lead-free relaxor BNT. The simulated spectrum (Fig.~\ref{fig:BNT}a)  
shows quantitative agreement with our experimental data from epitaxial BNT thin films (inset), reproducing the two key anomalies: a dielectric hump near \( T_d \approx 320 \)~K, followed by a broad plateau, and a subsequent peak at \( T_m \approx 640 \)~K. 
These results confirm the UniPero's applicability across different relaxor chemistries.
Furthermore, the simulation reveals that the macroscopic polarization decays smoothly upon heating, showing no abrupt changes at either $T_d$ or $T_m$. This again underscores the characteristic decoupling between the local dipole dynamics, which drive the dielectric response, and the evolution of long-range polar order captured by the macroscopic polarization.

Applying the same statistical analysis of ACFs to BNT at 300~K reveals that it also exhibits a dipolar nematic state over a broad temperature range. The \((\mu_j, \sigma_j)\) scatter plot (Fig.~\ref{fig:BNT}b) confirms that the local polarization dynamics are dominated by Type-III behavior (high $\mu$ and low $\sigma$), similar to the Pb-based relaxor. 
The arrow-cone visualization for a $3 \times 3 \times 3$ region reveals a similar absence of local parallel alignment. 
Although the system lacks polar clusters, the relative orientations between local polarizations remain statistically constrained due to the persistent polar directions of individual Type-III unit cells. 
This local disorder is accompanied by a loss of long-range polar order, as confirmed by the rapid decay of $C^p(r)$ to zero at 300~K (Fig.~S9).
However, a critical distinction emerges when examining the global orientational preference. Unlike the uniaxial orientational alignment observed in the Pb-based system (Fig.~\ref{fig:acf}f), the ensemble polarization sphere for BNT (Fig.~\ref{fig:BNT}d) reveals eight distinct, symmetry-related polar orientations. 
These results suggest that BNT adopts a ``multipolar nematic state" in the absence of long-range polar order, in which individual local polarizations fluctuate around one of eight symmetry-related preferred axes. At higher temperatures, the ensemble polarization sphere approaches isotropy (see Fig.~S12).

\subsection{Dipolar nematic state in Ba-based relaxors}
\begin{figure}[t]
\includegraphics[width=0.9\textwidth]{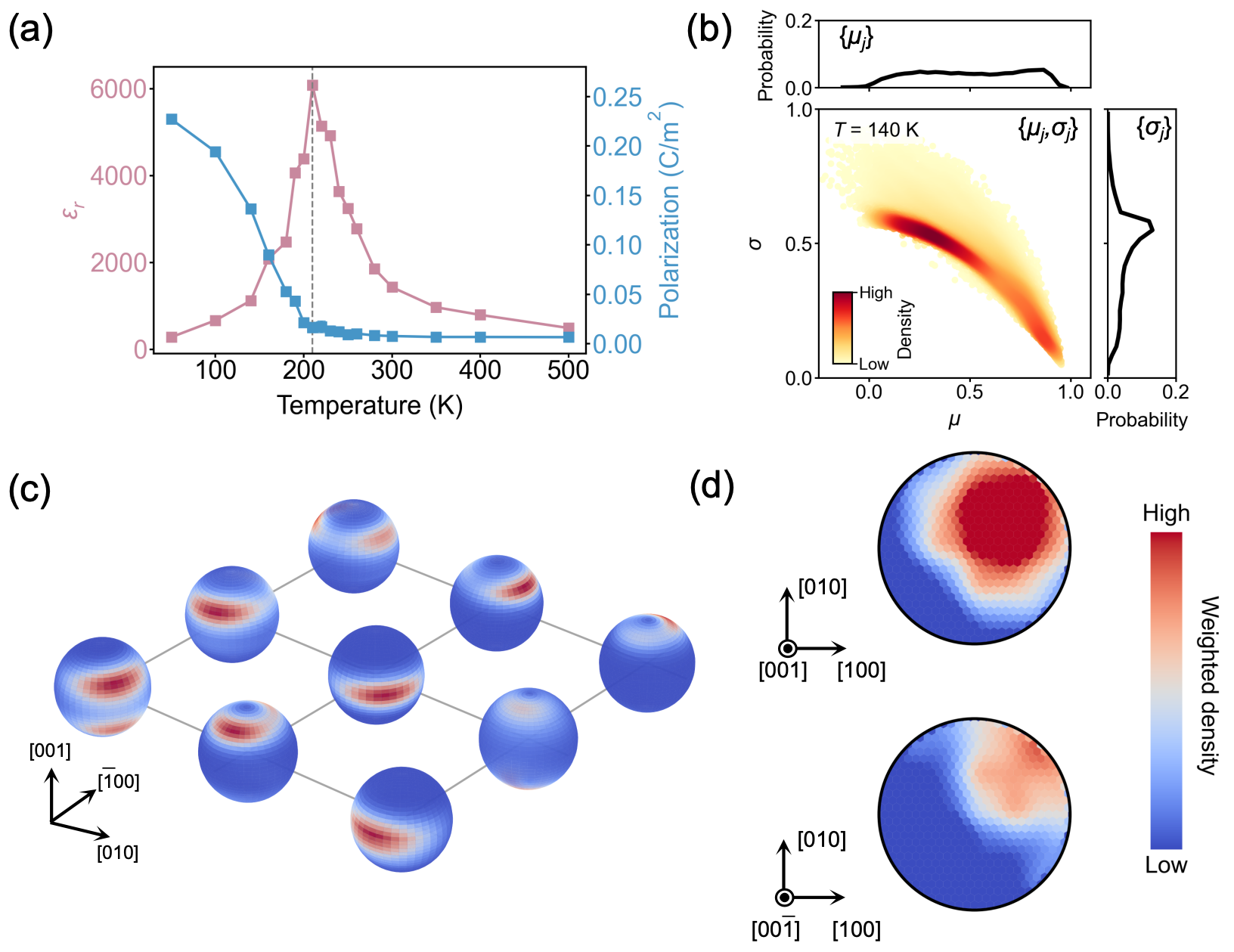}
\caption{\textbf{Dipolar nematic state in the Ba-based relaxor BaZr$_{0.3}$Ti$_{0.7}$O$_3$.}
\textbf{a}, Simulated dielectric permittivity and total polarization of BaZr$_{0.3}$Ti$_{0.7}$O$_3$ as functions of temperature, showing a broad dielectric peak at $T_\mathrm{m} \approx 210$~K.
\textbf{b}, Scatter plot of the $\{\mu_j$, $\sigma_j\}$ distributions at 140~K. The broad distribution of $\{\mu_j\}$ indicates a significant fraction of dynamic Type-I and Type-II cells exhibiting high orientational variance.
\textbf{c}, Local polarization trajectories  projected onto unit spheres for a representative $3 \times 3$ region, showing that dipoles fluctuate within confined zones rather than around a single axis.
\textbf{d}, Despite pronounced local dynamic disorder, the ensemble polarization sphere reveals a single dominant polar orientation, confirming the emergence of global orientational coherence characteristic of a dipolar nematic state.}
\label{fig:BZT}
\end{figure}

Finally, we investigate a Ba-based relaxor system, Ba(Zr$_{0.3}$Ti$_{0.7}$)O$_3$ (BZT). MD simulations using the UniPero model again reproduce the hallmark features of a relaxor. As shown in Figure~\ref{fig:BZT}a, the system exhibits a broad peak in dielectric permittivity with a maximum at $T_m \approx 210$~K, accompanied by a gradual and smooth reduction in total polarization with increasing temperature.
The ($\mu_j$, $\sigma_j$) plot at 140~K (below $T_m$), shown in Fig.~\ref{fig:BZT}b, reveals features distinct from the Pb- and Bi-based systems. The $\mu$-distribution is broad and nearly uniform from 0.2 to 0.8, while the $\sigma$-distribution is wide, peaking near $\sigma = 0.5$. This indicates that a substantial fraction of unit cells belong to Type-I and Type-II, which are characterized by high orientational variance, including stochastic reorientations.

Given this pronounced local disorder, the arrow-cone visualization used for Pb- and Bi-based relaxors is not suitable. Instead, we adopt the unit-sphere projection to visualize the polarization trajectory for each unit cell. As illustrated in Fig.~\ref{fig:BZT}c for a representative 3~$\times$~3 region, local polarizations fluctuate within confined zones of the unit sphere rather than processing around a single, well-defined axis (as in Type-III cells).
Remarkably, despite significant local multipolar behavior, a global orientational coherence emerges. As shown in Fig.~\ref{fig:BZT}d, the ensemble polarization sphere reveals a single dominant polar orientation across the system. This coexistence of weak local alignment and global orientational order suggests that BZT adopts a dipolar nematic state in the absence of long-range polar order ($T \gtrsim  220$~K, see $C^p(r)$ in Fig.~S10), which evolves into a ferroelectric nematic state upon further cooling as the macroscopic polarization becomes substantial (see Fig.~S13).

\section{Discussion}
Our comprehensive MD simulations, validated by their ability to reproduce key properties of relaxor ferroelectrics, uncover a dipolar nematic state as a unifying feature across the seemingly disparate Pb-, Bi-, and Ba-based relaxor systems. This state is characterized by long-range orientational correlations coexisting with pronounced local disorder, moving beyond classical PNR or PND models by demonstrating that orientational coherence can emerge without locally ordered clusters.

\begin{figure}[t]
\includegraphics[width=0.9\textwidth]{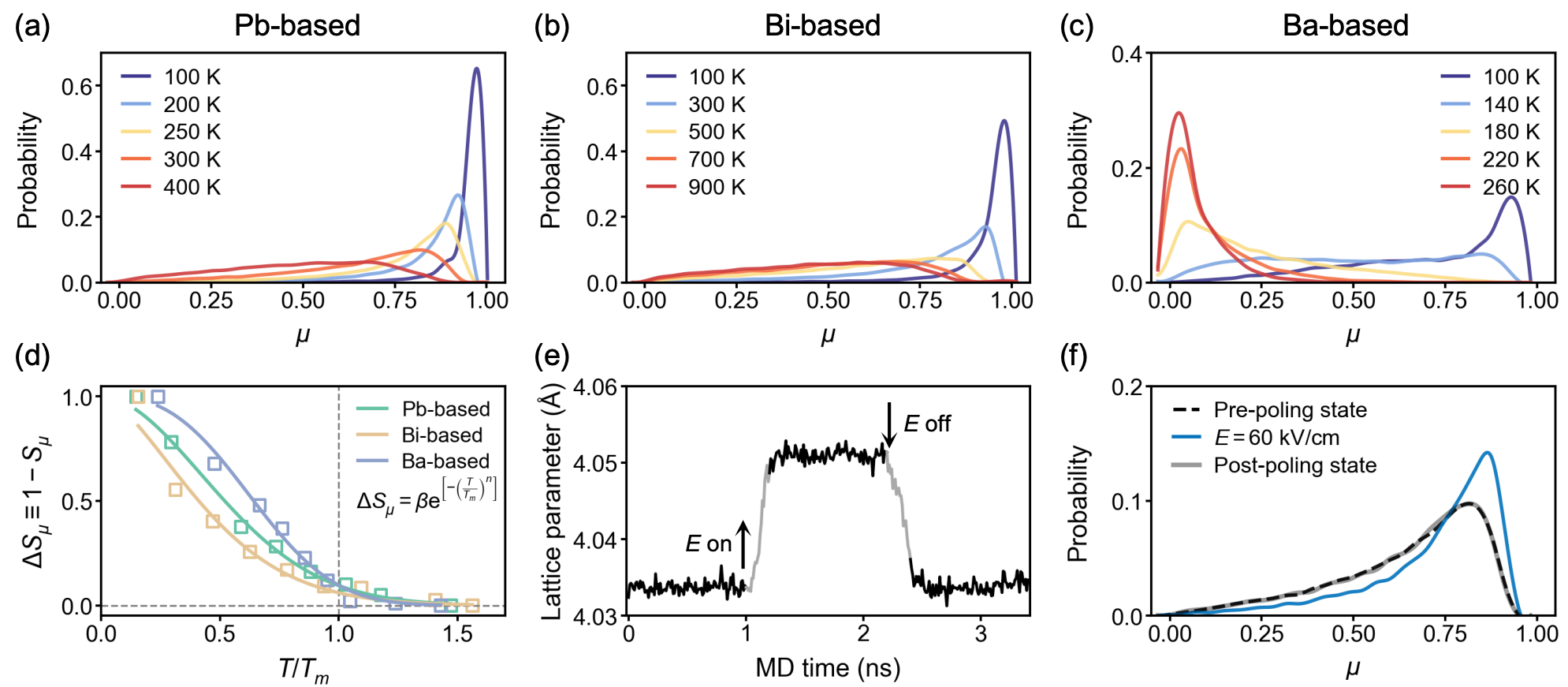}
\caption{
\textbf{Universal nematic ordering and its role in functional properties.}  
\textbf{a--c}, Evolution of the \( \mu \)-distribution upon heating for PIN-PMN-PT, BNT, and BZT, respectively. All three systems exhibit a universal trend where the peak shifts to lower \( \mu \) values and broadens, corresponding to a thermally driven conversion from Type-III to Type-I and Type-II unit cells.  
\textbf{d}, Universal order parameter (\( \Delta S_\mu = 1 - S_\mu \)), derived from the skewness of the \( \mu \)-distribution, plotted against normalized temperature (\( T/T_m \)). Data from all three chemically distinct relaxors collapse onto a single master curve that decays to zero as \( T \to T_m \), quantitatively linking the loss of nematic order to the macroscopic dielectric peak.  
\textbf{e}, Strain response of PIN-PMN-PT under a 60~kV/cm electric field, showing a giant and reversible piezoelectric effect with an effective $d_{33}=1270$~pm/V.  
\textbf{f}, Evolution of the \( \mu \)-distribution before applying and after removing the electric field. The complete recovery of the $\mu$-distribution demonstrates that the dipolar nematic state is robust and responsible for the structural memory underlying the reversible electromechanical response.
}
\label{fig:all}
\end{figure}

The temperature dependence of the $\mu$-distributions reinforces the universality of the dipolar nematic picture. As shown in Fig.~\ref{fig:all}a-c, all three systems display a similar evolution upon heating: the distribution's peak shifts from high $\mu$ toward zero and broadens. 
The $A$-site-driven systems, such as PIN-PMN-PT and BNT, where $A$-site has ferroelectrically active ions, evolve in an almost identical manner. The $B$-site-driven system BZT, despite its greater intrinsic disorder, follows the same underlying trend.

To quantify this universal thermal evolution, we propose the skewness of the $\mu$-distribution, $S_\mu$, as a robust order parameter for the dipolar nematic state. We use the quantity ($1 - S_\mu$), with $S_\mu$ normalized with respect to its high-temperature value, such that it behaves as a conventional order parameter, decaying from near unity in the low-temperature ordered state to zero in the high-temperature disordered state. The effectiveness of this approach is demonstrated in Fig.~\ref{fig:all}d, where the data from all three chemically distinct relaxors collapse onto a single, universal master curve when plotted against the normalized temperature, $T/T_m$. Crucially, this curve decays to zero as $T/T_m \approx 1$, explicitly linking the disappearance of microscopic nematic order to the macroscopic dielectric maximum. This universal trend is well-described by an empirical stretched exponential function, $\beta \exp(-(T/T_m)^n)$, where the exponent $n$ quantifies the diffuseness of the transition. 

The dipolar nematic model thus provides a natural framework for understanding the diffuse phase transition in relaxors. Unlike conventional ferroelectrics, where long-range polar order collapses abruptly at the Curie temperature, the transition here occurs in two stages. Upon heating, the system first loses macroscopic polarization but retains orientational correlations, entering the dipolar nematic state (Figs.~S11-S13). The broad dielectric peak at $T_m$ reflects a dynamic crossover, driven by the increasing population of thermally activated Type-I and Type-II cells (Fig.~\ref{fig:acf}a–c), which optimize the system's field response. This gradual evolution, captured by the decay of the universal order parameter ($1-S_\mu$), as shown in Fig.~\ref{fig:all}d, underlies the diffuse nature of the relaxor transition.

Furthermore, the dipolar nematic state, with its intrinsic spectrum of local dynamics, offers a clear physical basis for the frequency-dependent dielectric dispersion common in relaxors. Our classification of unit cells into persistent (Type-III), diffuse (Type-II), and stochastic (Type-I) types corresponds to a broad distribution of polarization relaxation times. The measured dielectric permittivity is thus a convolution of the responses from these distinct dynamic populations. At low frequencies, all cell types contribute, yielding high permittivity. As frequency increases, slower Type-III cells can no longer follow the field, leading to a reduced dielectric response. Frequency-dependent anomalies, such as $T_d$~in BNT (Fig.~\ref{fig:BNT}a), can therefore be interpreted as the freezing out of specific dynamic modes within the nematic ensemble, a behavior not easily explained by static PNR models.

Finally, we demonstrate that the $\mu$-distribution, and by extension the dipolar nematic state, acts as a form of ``structural memory" underlying the reversible giant piezoelectric response in relaxors. As shown in Fig.~\ref{fig:all}e, applying a 60 kV/cm electric field to the PIN-PMN-PT system induces a significant strain, corresponding to a giant effective piezoelectric coefficient of 1270 pm/V. 
This response originates from the orientational softness of the dipolar nematic state. Unlike conventional ferroelectrics, where dipoles are rigidly locked, local polarizations in the nematic state retain substantial rotational freedom within a correlated orientational framework. As a result, an external field can readily bias these fluctuations, promoting interconversion among unit cells with different dynamic behaviors (see the shift in the $\mu$-distribution in Fig.~\ref{fig:all}f). This enables large strain changes without the energetic penalty of disrupting strongly coupled domains.
Crucially, upon removal of the field, the system fully restores its original $\mu$-distribution (Fig.~\ref{fig:all}f), confirming that the dipolar nematic state is a robust ground state. 
Its ability to be reversibly perturbed and recovered underpins the exceptional reversible electromechanical properties.

\section{Conclusion}
In conclusion, our work establishes the dipolar nematic state as a unifying framework for understanding relaxor ferroelectrics across a broad range of chemistries, from lead-based to lead-free systems. This paradigm marks a fundamental shift from the traditional view of polar nanoregions and nanodomains. Rather than assuming clusters of aligned dipoles, we show that long-range coherence arises from statistical orientational correlations among spatially disordered local polarizations. The strength of this framework lies in its ability to quantitatively connect microscopic dipole dynamics with macroscopic properties, including diffuse phase transitions and giant piezoelectric responses, all derived from first principles. This is further exemplified by our universal order parameter, which collapses the thermal evolution of chemically distinct  relaxors onto a single master curve aligned with the dielectric maximum. Our findings not only provide a rigorous physical foundation for relaxor behavior but also offer a broadly applicable tool for understanding complex systems where order and disorder coexist.

\clearpage 
\section{Methods}
\subsection{Molecular dynamics}
All MD simulations are performed using a modified UniPero model.  
The original UniPero model is a deep neural network potential that incorporates a self-attention mechanism for latent-space communication between elemental and structural parameters~\cite{Zhang24p10}.  
The model was trained on a database constructed via a concurrent learning scheme~\cite{Zhang20p107206} and a modular development strategy~\cite{Wu23p144102} (see Ref.~\cite{Wu23p180804} for full details).  
However, while the self-attention layers improved feature mixing, they introduced excessive computational overhead for high-throughput MD simulations. By eliminating the attention layers and retraining on a slightly expanded dataset (19{,}773 configurations), we maintained accuracy while substantially enhancing computational efficiency. The streamlined UniPero, covering 14 metal elements, passed all regression tests and reproduced temperature-driven phase transitions across diverse ferroelectrics (see Figs.~S14-16). For transparency, the final training dataset, hyperparameters, and model are publicly available.  
The isobaric-isothermal (\(NPT\)) ensemble MD simulations are performed using the \texttt{LAMMPS} package~\cite{Plimpton95p1} with a time step of 2~fs.  
Temperature and pressure are controlled via the Nos\'e--Hoover thermostat and Parrinello--Rahman barostat, respectively. The three perovskite systems (PIN-PMN-PT, BNT, and BZT) are each modeled using a \(32 \times 32 \times 32\) supercell, containing 163{,}840 atoms.

\subsection{Polarization and static dielectric permittivity}
Given a configuration from MD simulations,  
the polarization for $AB$O$_3$ perovskite systems can be estimated using the following formula:
\begin{equation}
\bm{p}^m(t)=\frac{1}{V_{\rm uc}}\left[\frac{1}{8} \mathbf{Z}_{A}^* \sum_{i=1}^8 \mathbf{r}_{A, i}^m(t)+\mathbf{Z}_{B}^* \mathbf{r}_{B, i}^m(t)+\frac{1}{2} \mathbf{Z}_{\mathrm{O}}^* \sum_{i=1}^6 \mathbf{r}_{\mathrm{O}, i}^m(t)\right],
\end{equation}
where $\bm{p}^m(t)$ is the polarization of unit cell $m$ at time $t$, $V_{\rm uc}$ is the volume of the unit cell, $\mathbf{Z}_{A}^*$, $\mathbf{Z}_{B}^*$, and $\mathbf{Z}_{\mathrm{O}}^*$ are the average Born effective charges of the $A$ site, $B$ site, and O atoms (see Supplementary Sec.~III), and $\mathbf{r}_{A, i}^m(t)$, $\mathbf{r}_{B, i}^m(t)$, and $\mathbf{r}_{\mathrm{O}, i}^m(t)$ are the instantaneous atomic positions.
The static dielectric permittivity tensor, \(\epsilon_{ij}\), is computed from fluctuations of the total polarization ($\mathbf{P}$) using the Kubo formula~\cite{Liu16p094102,Boresch00p8743,Ponomareva07p227601,Ponomareva07p235403,Ponomareva08p012102} as  
\[
\epsilon_{ij} = \dfrac{\langle V \rangle}{\epsilon_0 k_B T} \left( \langle P_i P_j \rangle - \langle P_i \rangle \langle P_j \rangle \right),
\]  
where \(\epsilon_0\) is the vacuum permittivity, \(k_B\) is the Boltzmann constant, \(T\) is the temperature, and \(V\) is the supercell volume. The indices \(i\) and \(j\) denote Cartesian components, \(P_i\) represents the \(i\)-th component of the macroscopic polarization, and \(\langle \cdots \rangle\) denotes an ensemble average over equilibrium trajectories. The total dielectric response, \(\epsilon_r\), is evaluated as the modulus of the diagonal components,  
$\epsilon_r = \sqrt{\epsilon_{11}^2 + \epsilon_{22}^2 + \epsilon_{33}^2}$.

\subsection{Correlation functions}
Each equilibrium MD trajectory consists of a 1~ns simulation performed with a 2~fs integration timestep. Supercell configurations are recorded every 0.2 ps, resulting in a total of 5000 frames per trajectory.
The ensemble-averaged time-delayed local polarization correlation function, \( C^p(r, \tau) \), is defined as
\begin{equation}
C^p(r, \tau) = \frac{1}{N - \tau } \sum_{t = 1}^{N - \tau } \frac{1}{N^p(r)} \sum_i \sum_j \frac{\bm{p}_i(t)}{\|\bm{p}_i(t)\|} \cdot \frac{\bm{p}_j(t + \tau)}{\|\bm{p}_j(t + \tau)\|},
\end{equation}
where \( \bm{p}_i(t) \) is the instantaneous polarization vector of unit cell \( i \) at time \( t \), \( r \) denotes the spatial distance between unit cells \( i \) and \( j \), and \( \bm{r}_{ij} \) is the vector separating cells \( i \) and \( j \) such that \( \|\bm{r}_{ij}\| = r \). \( N \) is the total number of MD frames, \( \tau \) is the time delay (in frames) between two points in time, and \( t \) is the starting frame index, ranging from \( 1 \) to \( N - \tau \). \( N^p(r) \) denotes the number of unit cell pairs separated by distance \( r \).
The maximum number of samples available for a given delay \( \tau \) is therefore \( N - \tau \). Normalizing each polarization vector ensures that \( C^p(r, \tau) \) captures purely orientational correlations, independent of dipole magnitude.
The term ``ensemble-averaged" refers to averaging over all unit cell pairs at a given distance $r$ and over all time origins $t$, providing statistically robust measurements of correlation under equilibrium conditions.
A structurally analogous correlation function, \( C^d(r, \tau) \), is defined in the same manner, but with atomic displacement vectors substituted in place of local polarization vectors.

The ensemble-averaged local polarization autocorrelation function for a single unit cell $j$ corresponds to the special case \( r = 0 \), and is given by
\begin{equation}
A(\tau) = C^p(r = 0, \tau) = \frac{1}{N - \tau } \sum_{k = 1}^{N - \tau } A_k(\tau) = \frac{1}{N - \tau } \sum_{k = 1}^{N - \tau } \frac{\bm{p}_j(k)}{\|\bm{p}_j(k)\|} \cdot \frac{\bm{p}_j(k + \tau)}{\|\bm{p}_j(k + \tau)\|}.
\end{equation}
Here, \( A_k(\tau) \) represents the polarization autocorrelation function of unit cell \( j \) evaluated at time delay \( \tau \), starting from frame \( k \). The index \( k \) labels the starting point of each time interval. The evolution of \( k \) from 1 to \( N - \tau \) constitutes a sliding window analysis over the trajectory.

Instead of focusing solely on the ensemble-averaged autocorrelation \( A(\tau) \), we examine the full distribution of instantaneous autocorrelation values \( \{A_k(\tau = \Delta T)\} \) for each unit cell \( j \), where \( \Delta T \) is held fixed and \( k \in [1, N - \Delta T] \). Specifically, we set \( \Delta T = 1000 \) frames, corresponding to a physical duration of 200~ps.
From this distribution, we compute the mean \( \mu_j \) and standard deviation \( \sigma_j \) for each unit cell \( j \), which characterize the average persistence and temporal variability of the local polarization, respectively. Extending this analysis to all 32,768 unit cells in a $32\times32\times32$ supercell yields system-wide distributions of \( \{\mu_j\} \) and \( \{\sigma_j\} \) with $j \in [1, 32768]$, providing a statistical characterization of spatiotemporal polarization dynamics across the entire supercell.

\subsection{Diffuse neutron scattering calculations}
The inelastic neutron scattering cross section is given by \cite{Squires_2012,weadock2023nature}
\begin{equation}
\begin{gathered}
    \frac{d^2\sigma}{dE\,d\Omega} \propto \frac{k_i}{k_f} S(\bm{Q}, E)
\end{gathered}
\end{equation}
where \( k_i \) and \( k_f \) are the magnitudes of the incident and scattered neutron momenta, respectively, and \( S(\bm{Q}, E) \) is the dynamic structure factor (DSF). 
In this work, we omit constant prefactors that determine the overall magnitude of the cross section but are independent of the momentum transfer \( \bm{Q} \) and energy transfer \( E \). We also neglect the dependence on \( k_i \) and \( k_f \), and focus instead on the DSF, which encapsulates the intrinsic dynamical properties of the material and is independent of experimental parameters. 
Although the inelastic cross section is the direct experimental observable, it is the dynamic structure factor \( S(\bm{Q}, E) \) that encodes the microscopic dynamics of the system, and is therefore the quantity most commonly reported and analyzed in the literature.

In the classical approximation \cite{weadock2023nature}, the coherent DSF is
\begin{equation}
\begin{gathered}
    S(\bm{Q}, E ) = \left| \sum_i^N b_i \int e^{i\bm{Q} \cdot \bm{r}_i(t) - iEt/\hbar} \, dt \right|^2
    \label{eq:DSF}
\end{gathered}
\end{equation}
with \( i \) running over all atoms in the crystal, \( b_i \) the element-specific coherent scattering length, and \( \bm{r}_i(t) \) the trajectory of atom \( i \). The scattering lengths are known parameters \cite{sears1992neutron}, and the trajectories are calculated using classical MD; Eq.~\ref{eq:DSF} is then efficiently computed as a post-processing step.

In an experiment, there is finite energy resolution, so the elastic diffuse scattering intensity includes contributions from low-energy excitations. We define the diffuse scattering \( S(\bm{Q}) \) as a Gaussian-weighted average centered at \( E = 0 \). We use a typical Gaussian full width at half maximum (FWHM) energy resolution of \( E = 2 \) meV, $ S(\bm{Q}) = \int S(\bm{Q}, E) w(E) \, dE$ with $w(E) = \exp\left( -\dfrac{E^2}{2\sigma^2} \right)$, where \( \sigma \) being the standard deviation corresponding to FWHM = 2 meV.

The DSF was calculated using the \textsc{pynamic-structure-factor} code \cite{psf2023}. The spatial Fourier transform in Eq.~\ref{eq:DSF} is evaluated by explicitly summing over atoms; the time Fourier transform is performed using fast Fourier transforms. To achieve good statistics, the entire trajectory was divided into 20 blocks, each 50 ps in duration and containing 250 frames. 
The DSF was computed independently for each block and subsequently averaged over all 20 blocks. This setup yields a frequency resolution of 0.020\,THz (0.083\,meV) and a maximum frequency of 2.48\,THz (10.26\,meV). We verified that the results are robust with respect to both increased block length and potential correlations between blocks by repeating the analysis using non-consecutive segments.
The wave vector grid is defined with a spacing of \( 1/32 = 0.03125 \) reciprocal lattice units (r.l.u.), determined by the supercell dimensions of PIN--PMN--PT (\( 32 \times 32 \times 32 \)). The computed DSF was ``symmetrized'' by averaging over symmetry-equivalent wave vectors, consistent with standard practice in experimental data analysis. We confirmed that this symmetrization does not affect the conclusions of the study.

\clearpage

{\bf{Acknowledgments}} We acknowledge the support from National Natural Science Foundation of China (92370104) and Zhejiang Provincial Natural Science Foundation of China (LR25A040004). The computational resource is provided by Westlake HPC Center. T. C. S. was supported by the U.S. Department of Energy, Office of Basic Energy Sciences, Office of Science, under Contract No. DE-SC0024117. We also thank Professor Shiqing Deng for providing the experimental data.

{\bf{Author Contributions}} S.L. conceived the idea and designed the project. Y.L. performed the molecular dynamics simulations. T.S. computed the diffuse scattering patterns. Y.L. and S.L. analyzed the data and co-wrote the manuscript. All authors reviewed and commented on the manuscript.

{\bf{Competing Interests}} The authors declare no competing financial or non-financial interests.

{\bf{Data Availability}} The data that support the findings of this study are included in this article and are available from the corresponding author upon reasonable request.

\clearpage
\bibliography{SL}
\clearpage

\clearpage

\clearpage

\clearpage
\clearpage

\clearpage

\end{document}